# Violet emission from bulk Si prompted by surface plasmon polaritons

Ventsislav M. Lavchiev[1,3*], Valerii M. Mikoushkin[2*], & Gang Chen[1*]

[1] Shanghai Institute of Technical Physics, Chinese Academy of Sciences, Yu Tian Rd. 500, 200083 Shanghai, China

[2] Ioffe Institute, Polytechnicheskaya str. 26, 194021 Saint-Petersburg, Russia

[3] Institute for Microelectronics and Microsensors, Johannes Kepler University, Altenbergerstr. 69, 4040 Linz, Austria

*- These authors contributed equally to this work

**Silicon has been long known as a poor light emitter due to its indirect band gap and strong phonon-assisted decay of the excited states. Nevertheless, we have revealed efficient quasi-monochromatic photoluminescence at 368 nm from bulk silicon in the near-violet spectral range of the interband transition $SiE_{\Gamma1}$ even at room temperature. Optical and electron spectroscopy experiments showed a clear relation of the emission to a surface plasmon polariton (SPP) located on the SiO/Si interface. The presented results demonstrate that the unveiled luminescence is an effect of creation and decay of the SPP being a mixture of the surface plasmon, light, and the bulk interband transition. The SPP evolves in the silicon layer with a thickness of the order of the near-violet wave penetration depth (~80 nm) and is characterized by high excitation amplitude due to its hybrid nature. The effect seems to be easily applicable in developing near-violet light sources based on the existing Si-technology.**

Silicon technology has paved the way into the information age due to its ultrahigh integration capability and precise control of the circuit properties and geometry at the nanoscale. In particular, silicon CMOS- compatible photonics has been booming in recent years to meet the everlasting requirements from the industry for higher information speed and capacity, lower dissipation, and lower cost[1]. Despite the success in fabricating a variety of Si-based photonic devices - waveguides, modulators, switches, photodetectors - there still remains a major challenge to obtain an efficient light emitter which could be incorporated into the Si-based electronic and photonic devices. The obstacle is the low light emitting efficiency of Si caused by the indirect band gap or prevailing phonon-assisted electron transitions. To break this bottleneck, great efforts have been devoted to exploit surface and quantum confinement effects in various Si-based materials such as porous Si[2], silicon nanocrystals[3,4] and nanowires[5], germanium (Ge) quantum dots in silicon[6-8]. Er-doped Si was shown to yield light emission at the telecom wavelength due to transition involving atomic inner shell in Er[9]. Defects induced by material imperfection, doping, irradiation can also lead to strong luminescence[10-13]. Finally, mechanical strains arising in Ge/Si and GeSn/Si structures modify the indirect band gap thus leading to enhancement of light emission[14].

Recently, plasmonic effects have been applied to boost light emission from Si nanoclusters and Si-noble metal compounds[15]. High local electromagnetic fields generated by surface plasmons modify the absorption cross-section, radiative decay rate and quantum efficiency of Si nanocrystals[16,17]. The reduced surface plasmon polariton (SPP) wavelength allows for high Q-factor microcavities[18] and nanocavities[19], confining light below the diffraction limit (1)). However, a more straightforward way enabling desirable expansion of the wave-length range is to develop Si-compatible plasmonic materials beyond the noble metals[20-22]. Although the existence of SPP was predicted in heavily doped Si[23] and observed in transitional metal silicides[24,25] and Ge nanocrystals embedded in $SiO_2$ matrix[26], there have been no reports yet on plasmonic enhancement of light emission with these materials. Thus, the non-metal plasmonic approach remains a topical problem.

Here we report revealing the narrow-band near-violet photoluminescence (PL) from an ordinary Si wafer. The study with electron energy loss and Auger electron spectroscopies revealed existence of surface plasmon polariton at the SiO/Si interface. Furthermore, the energy of the SPP at the SiO/Si interface virtually coincides with the observed PL line. The temperature-dependent PL investigation has shown that the revealed luminescence is not related to the net electron-hole recombination, but it is assisted by the SPP. This result provides a novel but simple mechanism of light emission from bulk Si, which can be explored as an alternative approach to achieve a seamlessly integrated Si light emitter especially in the near-violet range.

**Photoluminescence of Si and Si/Ge samples in the near-violet spectral range**

Photoluminescence from two types of samples has been studied: Si(001) wafer and Si/Ge structure containing sparse submicron clusters of Ge. Both samples were kept under atmospheric conditions prior to inserting into a cryostat. A He-Cd laser with a wavelength $\lambda_{exc}$ = 325 nm ($h\nu$ = 3.81 eV) was used for excitation. Figure 1a demonstrates PL spectra of Si(001) obtained at temperatures T = 11°K (top panel) and T = 200°K (bottom panel).

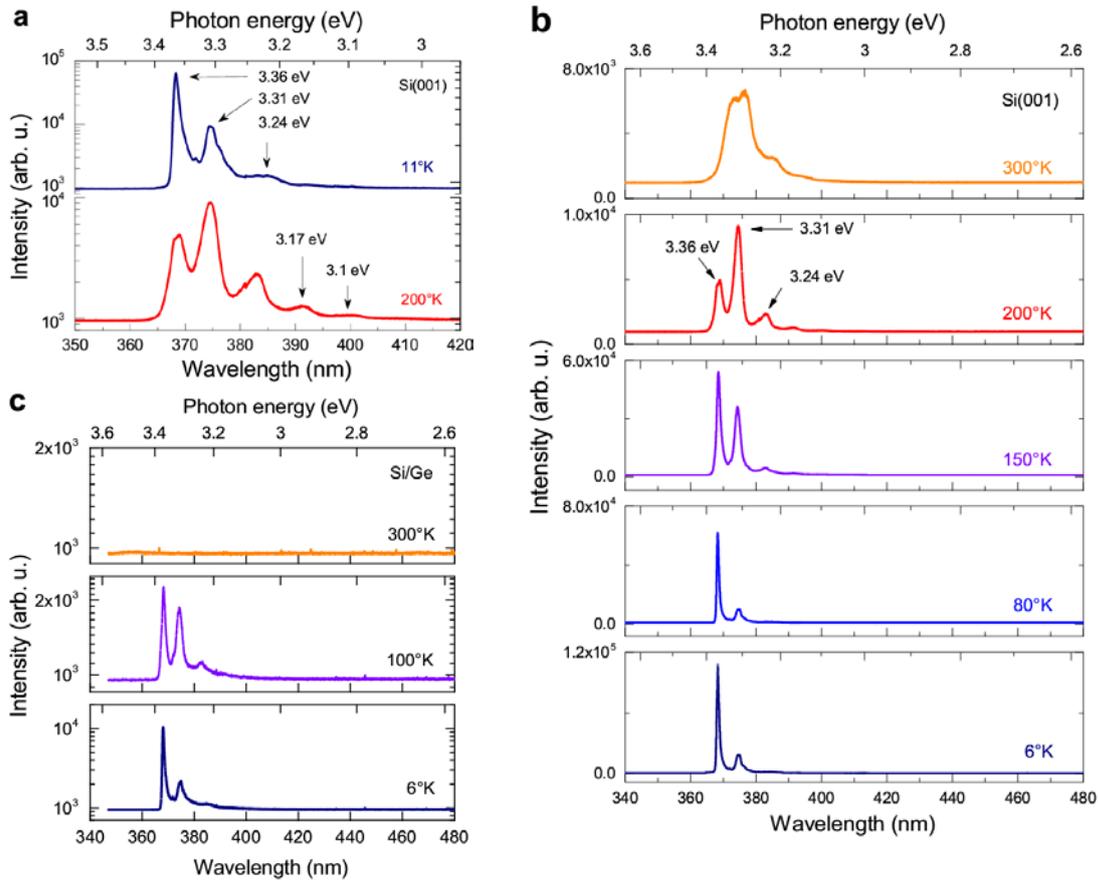

**Figure 1 | Photoluminescence spectra of Si(001) and Si/Ge under excitation by UV laser of $\lambda_{exc}$ = 325 nm.** (a) PL spectrum of Si(001) in the near-violet spectral range at T = 11°K and T=200°K. An intense, sharp peak at wavelengths $\lambda$ = 368 nm and a set of satellites at $\lambda$ = 375, 383, 392 nm are well resolved. (b) Temperature dependence of the PL signal of the Si(001) sample. (c) Temperature dependence of the PL signal from the Si/Ge sample.

The spectra reveal an intense quasi-monochromatic line at $\lambda = 368$ nm ($E_{ph} = 3.36$ eV) along with satellite peaks at $\lambda = 375, 383, 392, 400$ nm ($E_{ph} = 3.31, 3.24, 3.17, 3.1$ *eV*) arising at elevated temperatures (>150°K). The peak at 3.36 eV is recognized as the direct $E_{\Gamma1}$ transition in Si[27]. We identified the equidistant diminishing of the satellite photon energy to be related to energy losses of the primary photon (3.36 eV) due to excitation of one, two or more phonons of ~ 0.6 eV each. Similar PL spectra with essentially broader lines in the UV range of 340-440 nm were reported previously for porous Si[10,11] and nanocrystalline Si[12,13]. The luminescence has been assigned either to oxygen-related defects in porous Si or to the quantum confinement effect in nanocrystalline Si. Since we investigate perfect monocrystal Si(001), the PL lines cannot be linked to porosity or nanocrystallinity.

Figure 1b shows the temperature dynamics of the PL signal in a wide range, from liquid helium to room temperatures. The intense near-violet emission is observed over the entire temperature range. Noteworthy is that the emission is easily seen by naked eye as well, for which reason we call it "near-violet". The spectra also evidence the temperature stability of the PL peak position up to T ≈ 260°K (not shown here). The red shift of the PL lines between T = 6°K and T = 320°K does not exceed ~ 0.03 *eV*. In addition, the spectra does not show complete quenching of the luminescence usually observed due to strong phonon-assisted decay processes at elevated temperatures (typically above > 100°K). This behaviour contradicts the band theory of semiconductors and the Varshni relation implying a strong dependence between the energy band gap and the temperature[28-31]. The PL intensity, however, falls at room temperature by one order of magnitude in the Si(001) sample and even more in the Si/Ge sample which is characterized by a worse crystalline structure (Fig. 1c). Identity of the PL spectra of the both samples evidences a common nature of the observed luminescence related to Si surface.

Assuming an important role of the interface between Si and its oxides, the influence of the native oxide layer on the photoluminescence has been studied with a pair of Si(001) samples taken from the same wafer. One of them was left untreated while another one was treated by HF acid to remove the native oxide. Though, a thin native oxide layer emerged during sample transfer from the clean room to lab. Figure 2a compares signals from the two samples.

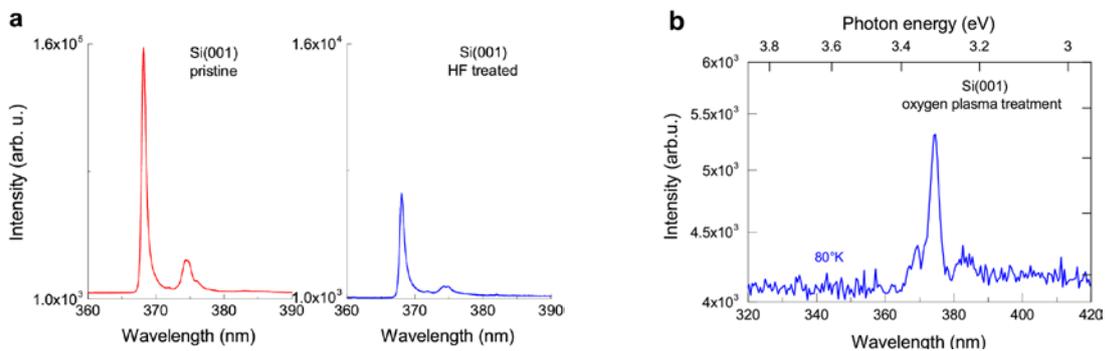

**Figure 2 | Influence of the native oxide layer on the photoluminescence.** (**a**) Comparison of photoluminescence at T = 6°K from Si(001) as received with that of Si(001) treated by HF. (**b**) PL intensity of Si(001) after treatment by oxygen plasma, The spectrum was measured at T = 80°K with a 213 nm laser.

The plot shows that the signal decreases by more than an order of magnitude due to oxide removal. At the same time, all the spectral features were reproducible despite essential diminishing of the oxide layer thickness. Still in a different experiment with a 213-nm laser, a Si(001) sample was treated with oxygen plasma in order to significantly enhance the oxidation extent of the Si surface (Fig. 2b). Again, a drastic drop of the peak intensity was observed. Both experiments demonstrated a crucial role of the oxide layer and its oxidation extent in the observed photoluminescence which disappeared after both the partial removal of the oxide layer and hard oxidation of the Si surface.

To explain the observed high intensity violet emission at $\lambda \sim 368$ nm, a research of the Si/Ge sample has been performed by Auger electron (AES) and electron energy loss (EELS) spectroscopies. The research was expected to enable us to check the hypothesis of PL enhancement in Si by SPP of the energy close to the observed PL line energy.

**Auger study of the Si,Ge/SiOx interfaces in the Si/Ge structure**

The sample surface was naturally oxidized before the experiment since the sample was stored in dry atmosphere. To determine the native oxide properties, the depth profile of the oxygen Auger O$KVV$ line intensity was measured in the course of Ar$^+$ etching of the Si-cap. Figure 3b demonstrates this profile. It shows that ion irradiation of the highest dose density ($Q_4$) removes a major part of the native oxide whose thickness is known to be $\sim 2$ nm[32] though depending on the material imperfection and experimental conditions[33]. Indeed, the oxide thickness exceeded the known mean free path of the Auger O$KVV$ electron in SiO$_2$ ($\lambda_O = 1.65$ nm[34]) after removal of a small part of the Si-cap by irradiation with a dose $Q_3$. Hence, the Si burying layer and the top of Ge clusters are dipped into dielectric silicon oxide medium, which is a traditional condition for the SPP existence. The dose density $Q_3$ was chosen to be the main measuring point in the EELS experiment.

The SPP energy strictly depends on both the chemical state of the surface atoms and permittivity of the interfacing dielectric media. Therefore, the oxidation extent of Si and Ge atoms was probed by AES whose basic advantage in this work over commonly used photoelectron spectroscopy is that AES provides information directly from the sample area of the EELS experiment, which is defined by the electron beam spot. Figures 3c,d demonstrate Si$KLL$ and Ge$LMM$ Auger spectra of the pristine sample (Q $\sim$ 0) and of the same sample after removing several atomic layers of the cap by Ar$^+$ ions (Q = $Q_3$). The Auger energies for Si, Ge and their dioxides[35] are indicated in the figure, as well as the peaks corresponding to the excitation of a bulk plasmon by Auger electrons ($\hbar\omega_B$).

The Si$KLL$ spectra reveal a dominating contribution of the SiO$_x$ phase throughout the layer thickness which is in between the mean free paths of Si$KLL$ Auger electron in SiO$_2$ ($\lambda_{Si} \sim 5.4$ nm[34]) and in Si ($\lambda_{Si} = 3.6$ nm[36]). Some contribution of the unoxidized Si is also seen. Taking into account the SiO$_x$ peak position being closer to the Si line, one should assume an average oxygen stoichiometry of $x \sim< 1$. Figure 3c also shows that only the upper part of the cap layer is oxidized to a higher extent since the SiO$_2$ contribution essentially diminishes after ion sputtering of several atomic layers. The Ge spectrum (Fig. 3d) evidences practically unoxidized state of Ge clusters. Thus the AES study concluded that the unoxidized Si-burying layer and the top of Ge clusters were initially covered by SiO$_2$-SiO$_x$ insulator medium $\sim$2-3 nm in average thickness. This condition seems to be favorable for the existence of SPPs on the SiO$_x$/Si,Ge interfaces.

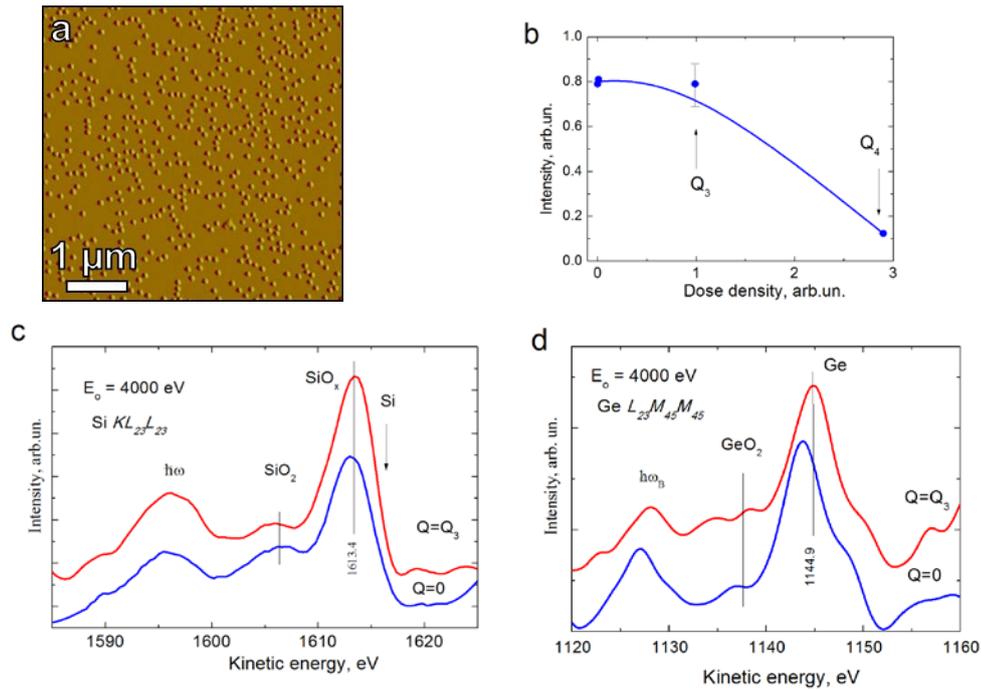

**Figure 3** | **Characterization of the elemental and chemical composition of the Si/Ge structure.** (a) SEM image of the Ge/Si sample surface. (b) Depth profile of the Auger O*KVV* line intensity determined in the course of Ar$^+$ etching of the Si-cap in the Si/Ge structure. (c) Si*KLL* Auger spectra of the Si/Ge structure measured before Ar$^+$ etching (Q = 0) and after removing several layers of the SiO$_2$ cap (Q = Q$_3$). (b) Ge*LMM* Auger spectra of the Si/Ge sample measured prior to Ar$^+$ etching (Q = 0) and after removing several layers of the SiO$_2$ cap (Q = Q$_3$).

**Surface plasmon polariton on the "inverse" SiO/Si interface studied by EELS**

Energy losses of the electrons recoiled on the SiO/Si,Ge interfaces, which could be associated with SPPs, were studied by the EELS method. Figure 4a shows EELS spectra in a wide loss energy range before (top) and after (bottom) subtracting the Shirley background. The most prominent broad peak (ΔE ~ 16÷25 *eV*) contains contributions from bulk plasmons. Plasmon energies used for the identification were taken from the literature[37-41]. The SiO-phase bulk plasmon energy crucial for this research was unknown. However, its value must be between the energies for Si and SiO$_2$ phases and close to the main peak (21 eV) since the SiO phase is dominant according to the AES data. Decomposition of the spectrum into constituents gave the exact value $\hbar\omega_B$(SiO) = 20.0 ± 0.2 eV.

Figure 4b shows the narrow low energy loss peak at ~ 3.5 eV in a larger scale. Second derivatives of the spectra revealed three to four contributions, three of which are visible even in the initial spectrum. The feature at ~ 6.5 eV was identified as the surface plasmon (6.5 - 7.5 eV) generated at the Si/SiO$_2$ interface[41,42]. The physical origin of this plasmon is considered to be the same as that for a "metal/insulator" interface, namely, the interaction of the bulk metal plasmon with the insulator electric field[43,44]. Si behaves here as a metal. Removal of the SiO$_2$ layer by Ar$^+$ etching results in complete disappearance of this contribution, thus reliably confirming the identification and applicability of a "metal/insulator" model to the surface plasmon in the Si/Si-oxide system.

The low-energy 3.6 eV peak resembles the corresponding spectra obtained previously for Si-oxide surfaces by the low-energy high-resolution EELS[42,45]. Two broad contributions in the vicinity of 3.5 and 5.3 eV were observed. They were explained previously by interband transitions in pure silicon. Using the positions of the lines in doubly differentiated spectra as starting values, the low-energy peak of the EELS spectra was decomposed into different contributions. Figure 4c presents the result of the decomposition of the initial (unirradiated) spectrum. Only decompositions into four contributions were acceptable. Among the contributions there were the known $Si/SiO_2$ surface plasmon and the lines at 3.4 and 4.8 eV that are close to the lines (3.5 and 5.3 eV) previously observed in the Si/Si-oxide system[42,45]. Minor contribution at 2.2 eV was also distinguished in this analysis in addition to the previously revealed lines. The loss energies and identification of the constituent lines from different EELS spectra are given in Table 1 and Fig 4b,c.

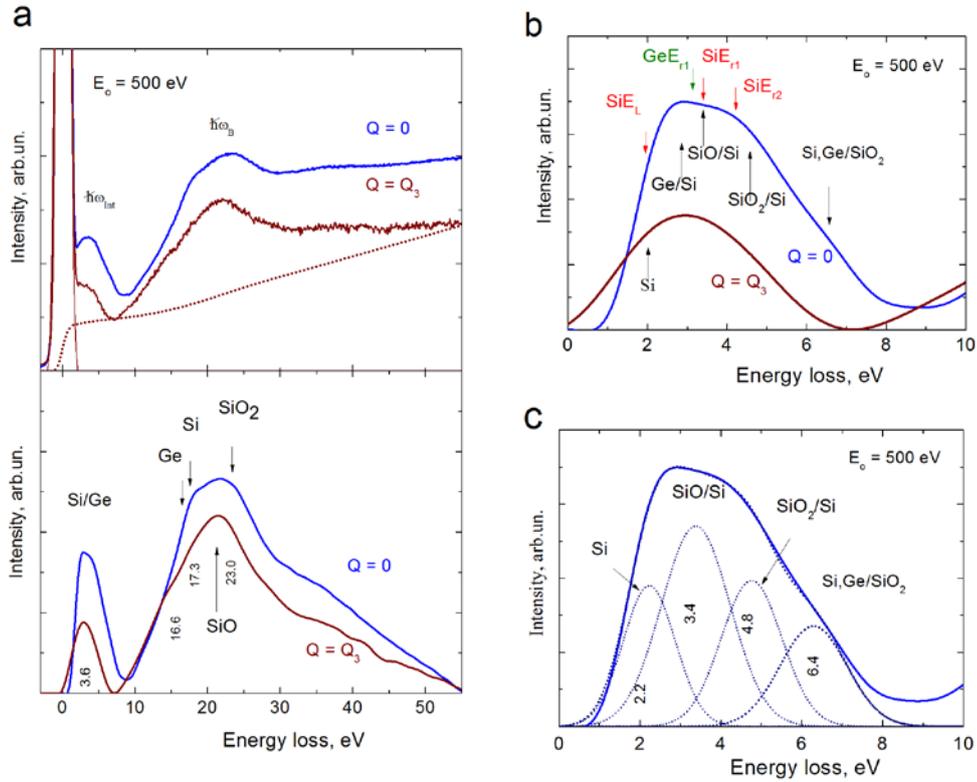

**Figure 4 | EELS spectra of the Si/Ge structure measured with $E_0$ = 500 eV electron beam before (Q = 0) and after (Q = $Q_3$) $Ar^+$ ion etching resulting in removal of several layers of the $SiO_2$ cap.** (a) EELS spectra before (top) and after (bottom) subtracting the background of multiply scattered electrons (dotted line) and the peak of the elastically scattered electrons. (b) Low-energy (~3.5 eV) peak in larger scale with the energy positions of the interband transitions and surface plasmons estimated. (c) Decomposed EELS spectrum of the unirradiated film.

Contrary to previous studies, the revealed lines were attributed not only to interband transitions but also to SPPs tied to interband transitions. To proof this identification, surface plasmon energies ($\hbar\omega_S$) for a set of interfaces were estimated in the frame of the Drude model applied to a metal/dielectric interface[43,44]:

$$\hbar\omega_S = \frac{\hbar\omega_B}{\sqrt{\varepsilon+1}} \quad (1)$$

where $\hbar\omega_B$ is the plasmon energy in the bulk of a metal, $\varepsilon$ is the permittivity of the dielectric medium above the surface (interface).

We used this approach for Si,Ge/Si-oxide interfaces first in the traditional way when semiconductor is considered as a metal. Instead of the static dielectric constant, the real part of the dielectric function $\varepsilon_1(\omega)$ at $\omega \sim \omega_S$ was used[46-49]. Table 1 shows that the surface plasmon energies exceed 5 eV even in the case of highest permittivity ($\varepsilon_1 = 9$) in slightly oxidized silicon (SiO$_{0.5}$). Therefore the lines with lower energies can be hardly explained in the frame of the traditional approach otherwise than by interband transitions[42,45].

**Table 1 | Surface plasmon energies (estimated and experimental) and their natural widths for interfaces of different materials in comparison with the band transition energies in Si and Ge.**

| A/B* | $\hbar\omega_S$ estim. eV | $\hbar\omega_S / \Gamma$ exp. eV | $\hbar\omega_B$ exp. eV | $\varepsilon_1(\omega)$ | SiE$_L$ eV | GeE$_{\Gamma2}$ eV | SiE$_{\Gamma1}$ eV | SiE$_{\Gamma2}$ eV |
|---|---|---|---|---|---|---|---|---|
| Si | - | 1.8-2.0 / 0 | - | | 2.0 | | | |
| Ge/Si | 2.9-3.2 | 3.2-3.4 / 0.8 | 16.0 | 25-30 | | 3.14 | 3.4 | |
| SiO/Si | 3.3-3.6 | 3.2-3.4 / 0.8 | 20 | 30-35 | | 3.14 | 3.4 | |
| SiO$_2$/Si | - | 4.8-4.9 / 0 | 23 | 10 | | | | 4.2 |
| Ge/SiO$_{0.5}$ | 5.1 | 4.8-4.9 / 0 | 16.0 | 9 | | | | 4.2 |
| Si/SiO$_{0.5}$ | 5.4 | 4.8-4.9 / 0 | 17.0 | 9 | | | | 4.2 |
| Ge/SiO | 6.3 | 6.4-7.0 / 1.5 | 16.0 | 5.5 | | | | |
| Si/SiO | 6.7 | 6.4-7.0 / 1.5 | 17.0 | 5.5 | | | | |
| Ge/SiO$_2$ | 7.2 | 6.4-7.0 / 1.5 | 16.0 | 3.9 | | | | |
| Si/SiO$_2$ | 7.7 | 6.4-7.0 / 1.5 | 17.0 | 3.9 | | | | |

\* Bulk plasmon $\hbar\omega_B$ is generated in layer A. Layer B plays a role of a dielectric in the A/B system. $\varepsilon_1(\omega)$ is the real part of the dielectric function in B at $\omega \sim \omega_S$.

The plasmon nature of some of these low-energy lines becomes apparent when the studied system is considered as an inverse SiO/Si system where the role of a metallic plasmon is played by the bulk plasmon in SiO semi-insulator and the role of dielectric is played by Si. Indeed, Si behaves as a dielectric since the real part of the Si-dielectric function reaches a value ($\varepsilon_1 \sim 40$) much higher than the modulus of the imaginary part ($\varepsilon_2 \ll \varepsilon_1$) in the range below strong interband transition ($\omega <$ SiE$_{\Gamma1}$ = 3.4 eV). The estimated value of SPP energy (Table 1) for the inverse SiO/Si system (3.3 - 3.6) almost fully coincides with the experimental value (3.2 - 3.4) of the main line in the low energy loss spectrum, which was previously identified as the interband transition (SiE$_{\Gamma1}$ = 3.4 eV) in the bulk Si[42,45]. We identified this energy loss as an SPP strongly tied with the interband transition SiE$_{\Gamma1}$. A set of arguments may be adduced in addition to the coincidence of the experimental line (~3.4 eV) with the estimated surface plasmon energy.

i) The PL line under consideration (3.36 eV) tends to vanish, as shown above, after partial removal of the SiO layer by HF acid or after oxygen plasma treatment enhancing the oxidation extent. This reference PL experiment proves that the ordinary interband transition in the bulk silicon cannot be solely responsible for the observed bright and extremely narrow (~5 nm) luminescence (3.36 eV). For instance, the pure interband transition SiE$_L$ (~2 eV) which has no entangled SPP does not produce any photoluminescence line of noticeable intensity.

ii) There is no temperature dependence of the 3.36 eV line in the PL spectrum typical of interband transitions in semiconductors.

iii) The natural width of the 3.4 eV loss energy line determined by subtracting the instrumental function proved to be very large ($\Gamma \sim 0.8$ eV), which assumes the lifetime as short as ~$10^{-15}$ s. On the contrary, the natural width of the ~ 2 eV loss line is at least one order of magnitude less and this line can be assigned only to the interband transition (SiE$_L$).

Therefore we associate the near-violet line with SPP in SiO/Si system. Small area of the Ge/Si interface does not play an essential role, as it was proved in the PL experiment with pure silicon discussed above. Though, the Ge/Si interface is expected to play the analogous role in similar luminescence at $hv = 3.14$ eV (Table 1). High brightness of the SPP line should be a result of creation and decay of the super SPP state which is a mixture of the ordinary SPP (surface plasmon plus light) and the bulk interband transition. This state evolves in the bulk Si layer within the penetration depth (~80 nm) of the entangled light and is characterized by large excitation amplitude due to its hybrid nature. Efficient excitation of the SPP by electron impact may be caused mainly by a swarm of secondary electrons due to the plasmonic constituent of the SPP. Photoexcitation develops as a result of the interband absorption followed by Stokes scattering towards the SiE$_{\Gamma 1} = 3.4$ eV point where the SPP is generated and decays with high rate light emission preventing further Stokes scattering to lower excitation energies. As a result, instead of the low-intensity emission in a wide spectral range in the course of Stokes scattering, an intense quasi-monochromatic light emission (3.36 eV) occurs.

**Conclusions**

Thus, extremely intense quasi-monochromatic photoluminescence ($\lambda = 368$ nm, $hv = 3.36$ eV) from bulk silicon has been revealed in the near-violet spectral range of the interband transition (SiE$_{\Gamma 1} = 3.4$ eV) under He-Cd laser irradiation ($\lambda = 3325$ nm, $hv = 3.815$ eV). A drastic decrease of the revealed PL line after partial removal of the native oxide by HF acid pointed out the crucial role of the interface between silicon and its oxide which was shown by AES to contain a thick SiO sublayer. EELS experiments revealed the energy loss line (3.2 – 3.4 eV) having a plasmonic character and virtually coinciding in energy with the PL line (3.36 eV). In turn, the electron energy loss proved to be very close to the estimate of the surface plasmon energy at the SiO/Si interface which was analysed in the frame of the Drude model applied to an "inverse" metal/dielectric interface where the role of a metallic plasmon was played by the bulk plasmon in SiO semi-insulator, while the role of dielectric was played by silicon. The collection of these facts led us to the conclusion that the quasi-monochromatic near-violet photoluminescence is a result of creation and decay of the surface plasmon polariton (SPP) being a mixture of the surface plasmon, light, and an additional contribution, namely, the bulk interband transition (SiE$_{\Gamma 1} = 3.4$ eV). The super SPP evolves in the bulk silicon layer with the thickness of the order of the near-violet wave penetration depth (~80 nm) and is characterized by high excitation amplitude due to its hybrid nature. Efficient excitation of the SPP by electron impact may be caused mainly by a swarm of secondary electrons due to the plasmonic constituent of the SPP. The revealed effect seems to be easily applicable in developing the near-violet light sources on the basis of the existing silicon technology.

**Methods and materials**

Two types of samples have been studied in a set of different experiments. The first ones were undoped Si(001) monocrystals obtained from wafers purchased at University Wafers and MTI Co. Samples of the second type were Si/Ge structures containing self-assembled Ge clusters grown by molecular beam epitaxy on Si(001) substrates[50]. The dome-type Ge clusters were ~15 nm in height and ~100 nm in diameter. Therefore, size-confinement effects were not expected in these structures. These samples were studied at the beginning of the research for searching the surface plasmon polariton (SPP) on the Si/Ge interface. However, this SPP was not distinguished on the background of the close in energy and extremely intense SPP related to the Si surface which exceeded 90% of the sample surface of the Si/Ge structures. This conclusion was confirmed by experiments with samples of the both types, which revealed virtually identical PL spectra connected with the SiO/Si interface as shown above. Both samples were kept under the atmospheric conditions prior to inserting them into a helium cryostat for the low-temperature measurements.

To clarify the role of Si-oxide, monocrystalline Si(001) samples were treated by HF (hydrofluoric) acid and oxygen plasma. The acid treatment removed the oxide layer, though a thin native oxide layer emerged in transferring the sample from the clean room to the PL lab. The plasma treatment significantly enhanced the oxidation extent of the Si surface. These experiments were considered to be the reference ones. They proved the crucial role of the Si-oxide layer in the Si-bulk light emission. In addition, observation of the light emission from the sample with extremely thin oxide layer emerged after the acid treatment rejected the assumption that the origin of this light is connected with the ordinary SPP arising from Si covered by a thick dielectric layer.

The photoluminescent measurements have been performed on two different setups to probe the reproducibility of the experimental data. The first setup included a He-Cd laser of a wavelength $\lambda_{exc}$ = 325 nm ($h\nu$ = 3.81 eV), a grating spectrometer Jobin Yvon Triax 550, and a Si CCD detector cooled with liquid-nitrogen. The other setup included a diode laser of $\lambda_{exc}$ = 213 nm, a grating spectrometer Idea Optics PG2000 Pro with a linear electrically-cooled Si CCD detector. While both setups reproduced the results, the first one offered much lower-noise spectra with a higher resolution. Typical instrumental resolution of the Triax 550 monochromator was 0.01 nm.

The EELS and AES experiments were carried out using electron spectrometer LHS-11 (Leybold AG) with ultra-high basic vacuum (P < 2 × $10^{-10}$ Torr) in the analytical chamber[51]. The chamber was equipped with a hemispherical energy analyser oriented normally to the sample surface. The EELS and AES spectra were measured under an electron beam targeting the sample at angle θ = 45°. For the EELS and AES experiments, a Si/Ge sample with Ge nanoclusters (6ML) was chosen which enabled studying interfaces of both Si and Ge with silicon oxide layers. The clusters occupied a minor part (~10%) of the sample surface and were purely dome-shaped, up to 150 nm in diameter and 10-12 nm in height. The structure had a thin protective Si cap layer whose thickness was designed to be as thin as possible (~2 nm) to enable using EELS and AES methods known as surface sensitive.

Germanium Ge *LMM*, silicon Si *LMM*, carbon C *KLL* and oxygen O *KLL* AES spectra were measured to control both the elemental composition of the near-surface layer of the sample and chemical states of the main elements (Ge, Si) in the studied structure. EEL spectra provided information on collective oscillations (plasmons) generated by primary electrons in the bulk of Ge-clusters and Si-burying layers. Excitation of interface (surface) plasmons was also expected. The energy of the primary electrons was chosen from the range of $E_e = 500 \div 1500$ eV in the EELS experiments and $E_e = 1500 \div 4000$ eV in the AES experiments. The background of the elastically and multiply scattered primary electrons in the EELS spectra was subtracted via the Shirley procedure. The information depth ($d$) in the EELS experiment a little greater than the electron mean free path $\lambda$ ($d = 1.2 \lambda$), taking into account that the incident and detection angles were 45° and 90°. The depths were $d = 2$ and 5 nm in $SiO_2$ medium for primary electron energies $E_0 = 500$ and 1500 eV, correspondingly. The data show that the 500-eV electron beam allows probing mainly the SiO/Si,Ge interfaces without contributing large background from the underlying Si burying layer or Ge clusters. Therefore, the data obtained with lower impact electron energy are mainly discussed. Information depth $d$ of the electrons detected in all the experiments did not exceed 4-5 nm, which was more than the cap layer thickness ($h \sim 2$ nm).

To remove contaminations from the sample surface and to reduce the thickness of the oxidized cap layer, the sample was etched by an $Ar^+$ ion beam from ion gun IQE 12/38 (Leybold AG). The ion beam was scanned over the surface (8×8 mm) essentially exceeding the studied area, which provided uniform etching. The ion energy was $E_i = 1500$ eV. It was planned to use only minimal $Ar^+$ ion etching sufficient to remove surface contaminations and to make the residual cap layer thickness a bit less than the detected depth. The pristine surface undamaged by ions was also studied.

**Acknowledgments**

V. M. Lavchiev expresses gratitude to the Austrian government for the partial financial support under the COMET program and to Prof. Wolfgang Jantsch for fruitful discussions and experimental support. G. Chen thanks the Hundred Talents Program of Chinese Academy of Science, the National Natural Science Foundation of China (No. 61474130) for the partial support of this work.

**Author contribution**

V. M. L. performed the PL experiments, Fig.1,2, participated in the AES-EELS experiments and in the discussion of the mechanism; G.C. carried out the MBE growth, participated in the PL measurements and in the discussion of the mechanism. V. M. M. performed AES and EELS experiments, Fig.2, 3, Table1, suggested the "inverse metal/dielectric structure" and "SSPP".

**Additional information**

The authors declare no competing financial interests. Reprints and permissions information is available at www.nature.com/reprints. Correspondence and requests for materials should be addressed to V. M. L.